\documentclass[article]{revtex4}
\usepackage{amsmath}
\usepackage{amssymb}
\usepackage{graphicx}
\usepackage{indentfirst}
\usepackage{txfonts}
\usepackage{dcolumn}
\usepackage{bm}
 \usepackage{epstopdf}

\newcommand{\be}{\begin{equation}}
\newcommand{\ee}{\end{equation}}
\newcommand{\bea}{\begin{eqnarray}}
\newcommand{\eea}{\end{eqnarray}}

\begin{document}
\title{Superradiantly stable non-extremal Reissner-Nordstrom black holes}
\author{Jia-Hui Huang}
\email{huangjh@m.scnu.edu.cn}
\author{Zhan-Feng Mai}
\affiliation{Laboratory of Quantum Engineering and Quantum Materials,
School of Physics and Telecommunication Engineering,
South China Normal University, Guangzhou 510006,China}
\date{\today}

\begin{abstract}
The superradiant stability is investigated for non-extremal Reissner-Nordstrom black hole.
We use an algebraic method to demonstrate that all non-extremal Reissner-Nordstrom black holes are superradiantly
stable against a charged massive scalar perturbation. This improves the results  obtained
before for non-extremal Reissner-Nordstrom black holes.
\end{abstract}
\maketitle

The stability problem of black hole is an important
topic in black hole physics. Regge and Wheeler\cite{wheeler1957}
proved that the spherically symmetric Schwarzschild black hole is
stable under perturbations. The stability problems of rotating or
charged black holes are more complicated due to the significant
effect of superradiance. Superradiance effect can occur in both
classical and quantum scattering processes
\cite{Manogue1988,Greiner1985,Cardoso2004}. When a charged bosonic
wave is impinging on a charged rotating black hole, the wave reflected by the
event horizon will be amplified if the wave
frequency $\omega$ lies in the following superradiant regime
  \begin{equation}\label{suprad re}
   0<\omega < m\Omega  + e\Phi,
  \end{equation}
where $m$ and $e$ are the azimuthal
harmonic number and charge of the incoming charged wave, $\Omega$ is the angular
velocity of black hole horizon and $\Phi=Q/r_H$ is the electric
potential of the black hole
\cite{P1969,Ch1970,M1972,Ya1971,Bardeen1972,Bekenstein1973}. This means that when the incoming
wave is scattered, the wave extracts rotational energy from
rotating black hole and electronic energy from charged
black hole. According to the black hole bomb mechanism proposed by Press and
Teukolsky\cite{PTbomb}, if there is a mirror between the black hole horizon and space
infinity, the amplified wave can be reflected back and forth
between the mirror and the black hole and grows exponentially.
This leads to the superradiant instability of the black hole.

The superradiant mechanism has been studied by many authors for the
(in)stability problem of black holes\cite{fn2004,Dolan,cardoso,li,zpw,r2013,hod2010,herdeiro2013}. Recently, for a
Kerr black hole under massive scalar perturbation, Hod has proposed
a stronger stability regime than before\cite{hodkerr}. The extremal
and non-extremal charged Reissner-Nordstrom (RN) black hole has been proved to be stable
against charged massive perturbation\cite{hodrn}. Similarly, the analog of charged RN
black hole in string theory has also been proved to be stable under
a charged massive scalar perturbation\cite{liranrn}.

In fact, up to now, the non-extremal charged RN black hole is proved to be superradiantly stable when
the mass $M$ and charge $Q$ of the black hole satisfy $(Q/M)^2\leq 8/9$ \cite{hodrn}.
In this paper, we demonstrate that the \emph{all} non-extremal charged RN black hole is stable against a massive
charged scalar perturbation. We find that there
is no trapping well outside the black hole, which is separated from
the horizon by a potential barrier. As the result, there is not
bound states in the superradiant regime, which can lead to the
instability of the charged RN black hole.

The metric of the RN black hole (in natural unit $G=c=\hbar=1$) is
\begin{equation}
ds^2=-(1-\frac{2M}{r}+\frac{Q^2}{r^2})dt^2+\frac{1}{(1-\frac{2M}{r}+\frac{Q^2}{r^2})}dr^2+r^2(d\theta^2+sin^2\theta d\phi^2),
\end{equation}
where $M$ and $Q$ are the mass and electric charge of the black hole.
The dynamics of a charged massive scalar field perturbation $\Psi$ is governed by the Klein-Gordon equation
\begin{equation}
[({\nabla ^\nu } - iq{A^\nu })({\nabla _\nu } - iq{A_\nu }) - {\mu ^2}]\Psi  = 0,
\end{equation}
where $q$ and $\mu$ are the charge and the mass of the scalar field. $A_\nu=-\delta_\nu^0 Q/r$ is the vector
potential that describes the spherically symmetric electric field.
The solution of the above equation with definite spherical harmonic
eigenvalues can be written as
\begin{equation}
\Psi_{lm}(t,r,\theta,\phi)=R_{lm}(r)Y_{lm}(\theta,\phi)e^{-i\omega
t},
\end{equation}
where $Y_{lm}$ is the spherical harmonic function, $l$ is the spherical harmonic index, $ m$ is the azimuthal
harmonic index with $-l\leqslant m \leqslant l$ and $\omega$ is the energy of the mode.
 The radial Klein-Gordon equation obeyed by $R_{lm}$ (we denote $R_{lm}$ by $R$ in the following) is given by
\begin{equation}\label{radeq}
\Delta \frac{d}{dr}(\Delta \frac{dR}{dr})+UR=0,
\end{equation}
where $\Delta=r^2-2Mr+Q^2$, and
\begin{equation}
U=(\omega r^2-qQr)^2-\Delta[\mu^2r^2+l(l+1)].
\end{equation}
The inner and outer horizons of the black hole are
\begin{equation}
r_{\pm}=M\pm\sqrt{M^2-Q^2},
\end{equation}
and it is obvious that
\begin{eqnarray*}
r_{+}+r_{-}&=2M,~~~r_{+}r_{-}&=Q^2.
\end{eqnarray*}

In order to study the superradiance stability of the black hole
against the massive charged perturbation, the asymptotic solutions
of the radial wave equation near the horizon and at infinity will be
considered with proper boundary conditions. By defining
the tortoise coordinate $y$ by equation $\frac{dy}{dr}=\frac{r^2}{\Delta}$ and a new radial
function as ${\tilde R}=rR$. The radial wave equation \eqref{radeq} can be written as
\begin{equation}
\frac{d^2\tilde R}{dy^2}+\tilde U\tilde R=0,
\end{equation}
where
\begin{equation}
\tilde U=\frac{U}{r^4}-\frac{\Delta}{r^3}\frac{d}{dr}(\frac{\Delta}{r^2}).
\end{equation}
It is easy to obtain the asymptotic behavior of the new potential $\tilde U$ as
\begin{eqnarray}
\mathop {\lim }\limits_{r \to {r_ + }} \tilde U = \frac{(\omega r_{+}-qQ)^2}{r_{+}^2},~~~~~~
\mathop {\lim }\limits_{r \to \infty } \tilde U =\omega^2- {\mu ^2}.
\end{eqnarray}
The chosen boundary conditions are ingoing wave at the horizon $(y\to -\infty)$ and bound states
(exponentially decaying modes) at spatial infinity $(y\to +\infty)$.
Then the radial wave equation has the following asymptotic solutions
\begin{equation}
 \tilde R\thicksim\left\{ \begin{array}{l}
{e^{-i(\omega-\frac{qQ}{r_+})y}},\quad \,\,y\to -\infty ~(r\to r_+) \\
 {e^{-\sqrt{\mu^2-\omega^2}y}},\quad \,\,y\to +\infty ~(r \to
 +\infty).\\
 \end{array} \right.
\end{equation}
It is obvious when
\begin{equation}\label{bound}
\omega^2<\mu^2
\end{equation}
there is a bound state of the scalar field.

In the following, we prove that there is no trapping well outside the black
hole horizon when parameters of the scalar field and the black hole
 satisfying the bound state condition \eqref{bound} and superradiance condition of RN black hole,
 \bea\label{superrad rn}
 0<\omega < q\Phi_H=qQ/r_+.
 \eea
We define a new radial function $\phi$ by $\phi=\Delta^{\frac{1}{2}}R$ ,
then the radial equation \eqref{radeq} can be rewritten as
\begin{equation}
\frac{d^2\phi}{dr^2}+(\omega^2-V)\phi=0,
\end{equation}
where
\begin{equation}
V =\omega^2-\frac{U+M^2-Q^2}{\Delta^2}.
\end{equation}
In order to see if there exist a trapping potential outside the
horizon, we should analyze the shape of the effective potential $V$.
From the following asymptotic behavior of the potential $V$
\bea
V(r \to +\infty)&\to& \mu^2+\frac{2M\mu^2+2Qq\omega-4M\omega^2}{r}+o(\frac{1}{r^2}),\\
V(r\to r_+)&\to& -\infty,
\eea
we know there is at least one maximum for $V$ outside the event horizon.
It is easy to see that the asymptotic behavior of the derivative of $V$ is
\begin{equation}
V'=-\frac{2M\mu^2+2Qq\omega-4M\omega^2}{r^2}+o(\frac{1}{r^3}).
\end{equation}
We can prove the coefficient $2M\mu^2+2Qq\omega-4M\omega^2>0$ when $\omega$ satisfies
the superradiance and the bound state conditions.
Define a quadratic function $f$ for $\omega$
\begin{equation}
f(\omega)=-4M\omega^2+2Qq\omega+ 2M\mu^2.
\end{equation}
It is obviously that there are two zero points for $f$ with opposite sign
and the positive one is
\begin{equation}
\omega_{+}=\frac{Qq+\sqrt{Q^2q^2+8M^2\mu^2}}{4M}.
\end{equation}
To verify $f(\omega)>0$ when $\omega$ satisfies
the superradiance and the bound state conditions, we just need to prove $\omega<\omega_+$.\\
Case I : $\omega<\mu\leq qQ/r_{+}$\\
With an obvious relation $r_{+}>M$, we can get
\begin{equation}
\omega_{+}=\frac{qQ}{4M}+\sqrt{\frac{q^2Q^2}{16M^2}+\frac{\mu^2}{2}}>\frac{\mu r_{+}}{4r_{+}}+\sqrt{\frac{\mu^2r_{+}^2}{16r_{+}^2}+\frac{\mu^2}{2}}=\mu>\omega.
\end{equation}
Case II : $\omega<qQ/r_{+}<\mu$\\
We can also easily get
\begin{equation}
\omega_{+}=\frac{qQ}{4M}+\sqrt{\frac{q^2Q^2}{16M^2}+\frac{\mu^2}{2}}>\frac{qQ}{4r_{+}}+\sqrt{\frac{q^2Q^2}{16r_{+}^2}+\frac{q^2Q^2}{2r_{+}^2}}=qQ/r_{+}>\omega.
\end{equation}
 So when $\omega$ satisfies
the superradiance and the bound state conditions, $f(\omega)>0$. It implies that
\begin{equation}
V'(r \to \infty) \to 0^-.
\end{equation}
This means there is no potential well when $r\to +\infty$.
In the following, we will show that there is only one maximum outside the event horizon for $V$,
no trapping potential exists which is separated from
the horizon by a potential barrier and all non-extremal RN black holes are superradiantly stable.

 The explicit expression of the derivative of the effective potential is
\begin{eqnarray}\nonumber\label{vdaor}
V^{'}&=&-\frac{1}{\Delta^3}[(-4M\omega^2+2qQ\omega+2M\mu^2)r^4+[4Q^2\omega^2+4MQq\omega-4M^2\mu^2-2Q^2(q^2+\mu^2)+2l(l+1)]r^3\\\nonumber
&+&[-6Q^3q\omega+6MQ^2\mu^2-6Ml(l+1)]r^2+[2Q^4q^2-2Q^4\mu^2-4M^2+4Q^2\\
&+&2(2M^2+Q^2)l(l+1)]r+4M^3-4MQ^2-2MQ^2l(l+1)].
\end{eqnarray}
Defining a new variable $z=r-r_{-}$ is convenient for us to study the property of the effective potential.
Then equation \eqref{vdaor} can be written as
\begin{equation}\label{vdaoz}
V'(z)=\frac{-1}{\Delta^3}(az^4+bz^3+cz^2+dz+e),
\end{equation}
where
\begin{eqnarray}
a&=&-4M\omega^2+2qQ\omega+2M\mu^2,\\
c&=&12r_{-}\{-4r_{-}^2\omega^2+6r_{-}\omega qQ-[2q^2Q^2+(3r_{+}r_{-}+r_{-}^2)\mu^2]\}-3(r_+-r_-)l(l+1),\\
e&=&2r_{-}^2(r_{+}-r_{-})(\omega r_{-}-qQ)^2+\frac{1}{2}(r_{+}-r_{-})^3.
\end{eqnarray}
From the asymptotic behaviors of the effective potential at the inner and outer horizons and infinity,
we know that there are at least two roots for $V'(z)=0$ when $z>0$. If a trapping
 potential existed, there would be at least four positive roots for $V'(z)=0$. Next, we will demonstrate
 that it is impossible for equation $V'(z)=0$ to have four positive roots when the superradiance condition \eqref{superrad rn}
 and bound state condition \eqref{bound} are satisfied.

Because we just concern about the roots of  $V'(z)=0$, the numerator of equation \eqref{vdaoz} will be considered only.
We denote the roots of $V'(z)=0$ by $\{z_{1},z_{2},z_{3},z_{4}\}$ and $z_1, z_2$ are the two known positive roots($r_-<z_1<r_+,r_+<z_2<+\infty$).
According the Vieta theorem, we have the following relations for the roots
\begin{eqnarray}\label{twoji}
z_{1}z_{2}+z_{1}z_{3}+z_{1}z_{4}+z_{2}z_{3}+z_{2}z_{4}+z_{3}z_{4}=\frac{c}{a},\\\label{siji}
z_{1}z_{2}z_{3}z_{4}=\frac{e}{a}.
\end{eqnarray}
The coefficient $a(=f(\omega))$ is proved to be positive before.
Because $r_{+}>r_{-}$, it is also easy to see that
\begin{equation}
e>0.
\end{equation}
So from equation \eqref{siji}, we find that if $z_3, z_4$ are two real roots, they must be both positive
or both negative.

Taking the superradiance and bound state conditions into account,
we will use an algebraic method to prove $c<0$ for the full parameter space of the charged massive scalar perturbation and non-extremal RN black holes.
So $z_3, z_4$ can not be both positive, there is no trapping well outside the horizon and
the RN black hole is superradiantly stable. This is the main result of this paper.

The final term of $c$ is non-positive, so in order to prove $c<0$,
we just need to prove
\bea
12r_{-}\{-4r_{-}^2\omega^2+6r_{-}\omega qQ-[2q^2Q^2+(3r_{+}r_{-}+r_{-}^2)\mu^2]\}<0,
\eea
 i.e.
\bea
g(\omega)=-4r_{-}^2\omega^2+6 qQ r_{-}\omega -[2q^2Q^2+(3r_{+}r_{-}+r_{-}^2)\mu^2]<0.
\eea
Regarding $g(\omega)$ as a quadratic function of $\omega$, when the discriminant of $g(\omega)$ (denoting it by $\Delta'$)
\begin{equation}
\Delta'=4r^2_{-}[q^2Q^2-4(3r_{+}r_{-}+r^2_{-})\mu^2]<0,
\end{equation}
we have $g(\omega)<0$ because the coefficient of
$\omega^2$ is negative.

Below we discuss the case $\Delta'\geq0$.
According to the properties of quadratic function, when $\Delta'\geq0$ is satisfied, there are two positive roots of $g(\omega)$ which are denoted by $\omega_{1}$ and $\omega_{2}$ respectively and $\omega_{1}\leq\omega_{2}$. To demonstrate $g(\omega)<0$, we only need to demonstrate $0<\omega<\omega_{1}$ when
the superradiance and bound state conditions are satisfied. We do this for two possible cases. It is easy to get
\begin{equation}
\omega_{1}=\frac{3qQ-\sqrt{q^2Q^2-4(3r_{+}r_{-}+r^2_{-})\mu^2}}{4r_{-}}.
\end{equation}
According to $\Delta'\geq0$ and $r_{+}>r_{-}$, we can obtain
\begin{equation}
qQ>4\mu r_{-}.
\end{equation}
Case I : $\omega<\mu\leq qQ/r_{+}$,
\begin{equation}
\omega_{1}=\frac{3qQ-\sqrt{q^2Q^2-4(3r_{+}r_{-}+r^2_{-})\mu^2}}{4r_{-}}>\frac{qQ}{2r_{-}}>2\mu>\mu>\omega.
\end{equation}
Case II : $\omega<qQ/r_{+}<\mu$,\\
It is also easy to get
\begin{equation}
\mu^2r_{+}^2>q^2Q^2>16\mu^2r_{-}^2,
\end{equation}
so that
\begin{equation}
r_{+}>4r_{-}.
\end{equation}
Then we have
\begin{equation}
\omega_{1}=\frac{3qQ-\sqrt{q^2Q^2-4(3r_{+}r_{-}+r^2_{-})\mu^2}}{4r_{-}}>\frac{qQ}{2r_{-}}>\frac{qQ}{4r_{-}}>\frac{qQ}{r_{+}}>\omega.
\end{equation}
The proof completes.

In summary, we study the superradiant stability of non-extremal charged RN black holes against a charged massive scalar perturbation.
Using an algebraic method, we demonstrate analytically that when the superradiance condition\eqref{superrad rn} and bound state condition\eqref{bound} are satisfied by the scalar perturbation and black holes, there is no trapping well outside the event horizon, which is separated from
the horizon by a potential barrier. So we conclude that\emph{ all} the non-extremal charged RN black hole is superradiantly stable.

\textbf{Note added}: after this paper was completed, ref.\cite{hodrnfin} appeared which addresses the same
issue with a different method and gets the same conclusion.

\end{document}